Title: A Novel Methodology for Evaluating Positive Phase Blast Wave Loading Parameters Using High Speed Video

Authors: Caio Barbosa Amorim[1,2], Clare Knock[2], Dain George Farrimond[3,4], Rene Francisco Boschi Gonçalves[1]

[1] Instituto Tecnológico de Aeronáutica, Praça Marechal Eduardo Gomes, 50, Vila das Acácias, São José dos Campos, 12228-900, SP, Brasil

[2] Centre for Energetics Technology, Cranfield University, Defence Academy of the UK, Shrivenham, SN6 8LA, UK

[3] School of Mechanical, Aerospace & Civil Engineering, University of Sheffield, Mappin Street, Sheffield, S1 3JD, UK

[4] Blastech Ltd. The Innovation Centre, 217 Portobello, Sheffield, S1 4DP, UK

Abstract: Traditionally, the critical blast wave parameters used to characterize loading conditions are obtained through pressure gauge measurements. However, these instruments are costly, require careful calibration, provide discrete location measurements only, and must be deployed in hazardous environments. Recent events, such as the Beirut port explosion have demonstrated that video recordings, which provide time of arrival ($t_a$) versus distance data, offers valuable information for post-event blast analysis. However, methodologies capable of predicting key blast parameters, such as positive phase duration and impulse, using video data alone remain limited. This work proposes and validates a novel methodology to predict positive phase duration and impulse for spherical, non-cased, free air bursts of ideal explosives using $t_a$ data only. The proposed methodology was evaluated using experimental datasets from the literature for bulk and cartridge PE4, PE7, Composition B, and PETN. The positive phase duration and impulse models achieved, respectively, mean absolute percentage errors of 5.3% and 5.3%, maximum deviations of 20% and 9.4%, absolute biases of zero and 3.1%, and confidence interval coverages of 86% and 83%. The predicted results achieve remarkable comparison to all reported experimental data, verifying the ability to capture positive phase blast loading for high speed video; a step-change in explosive characterisation through full spatial and temporal primary shock characteristics.

## 1. Introduction

Detonation reactions may occur intentionally or accidentally in both military and civilian environments. Blast waves are one of the primary phenomena associated with these detonation reactions and are responsible for a significant proportion of the damage on infrastructure. Hence the blast wave's magnitude is a key parameter when designing structures to withstand such loads. When a detonation occurs in air, the rapid conversion of a solid explosive material into high-temperature, high pressure gases generates a shock wave that propagates outward from the source. This wave produces a sudden rise in pressure at a given point in space, followed by a gradual decay toward ambient conditions [1].

The overpressure-time profile associated with a blast wave is typically characterized by several key parameters, as shown in Fig. 1. The time of arrival ($t_a$) represents the time from the detonation to the arrival of the shock front at a given point, while the maximum overpressure ($P_{max}$) quantifies the pressure rise experienced at that location. This moment marks the beginning of the positive phase, ending when the overpressure returns to zero. The duration of this phase is referred to as the positive phase duration ($t_d$) and the integral of the overpressure over this period is defined as the positive phase impulse ($I_+$). These parameters are widely used in blast engineering to evaluate the damaging potential of explosions and to assess the vulnerability of structures subjected to blast loading [1]-[2].

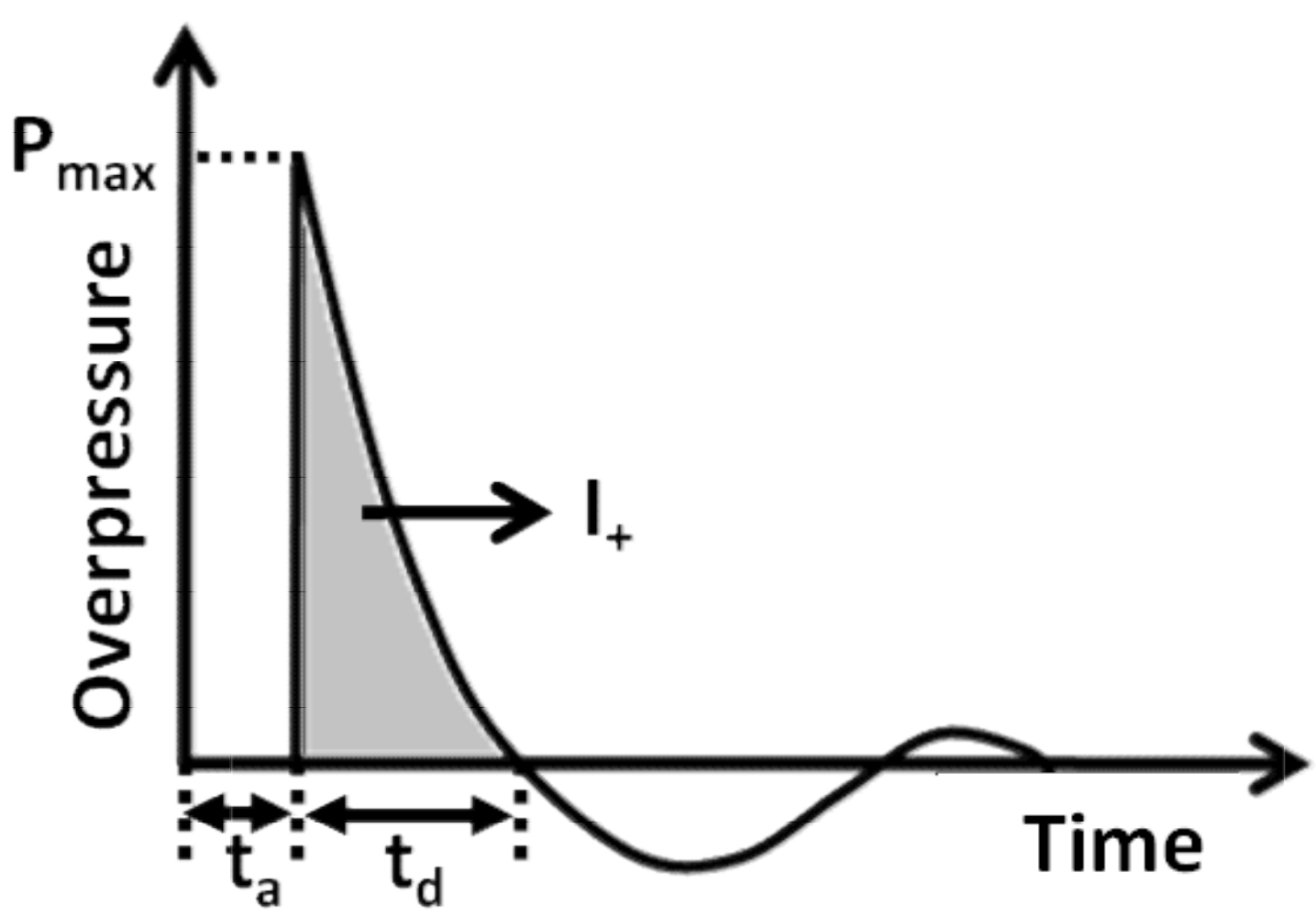

Fig. 1. Overpressure-time profile of a blast wave.

The experimental determination of blast wave parameters is traditionally performed using pressure gauges placed at specific discrete distances from an explosive charge. These instruments are designed to withstand extreme pressure and temperature conditions and must be carefully calibrated to ensure reliable measurements and thus are expensive to acquire and maintain. They must also be positioned relatively close to the explosion, exposing instrumentation and personnel to potentially hazardous conditions. These limitations may restrict the availability of high-quality experimental data, particularly for accidental explosions or in situations where the deployment of instrumentation is not feasible.

On the other hand, a high speed camera is a reliable and versatile instrument for measuring different parameters derived from blast events. It can be used to track the detonation velocity of an explosive as it propagates within the composition [3], the blast wave position as it expands away from the detonation point [4], measure the expanding fireball temperature and chemical species development [5], any casing fragment trajectories [6], and structural displacements under blast loading [7]. Without a high speed camera, each aforementioned application would probably require a specific instrument to acquire similar data.

High speed video recordings can be obtained from safe distances and are therefore less affected by the extreme environmental conditions near the explosion source. In addition to dedicated high speed cameras used in experimental setups, modern urban environments contain a large number of surveillance cameras and personal recording devices. As a result, explosive events are increasingly being documented through video footage.

The explosion that occurred at the Port of Beirut in 2020 highlighted the potential value of such recordings. In that case, numerous low-cost personal device videos captured from different locations were later analysed to predict the explosive yield [8]-[12]. Further investigations of pre- and post-satellite data resulted in the development of machine learning models to predict damage levels from the Beirut blast. The models were trained with on-the-ground structural damage information and blast simulation and resulted in high agreement levels with the case study data – emphasising the use of optical methods for understanding loading conditions from complex urban blast events [13]. These developments suggest that optical data provides an alternative source of information for full-field spatial blast analysis when conventional measurements are unavailable.

These investigations normally use the Kingery and Bulmash (KB) semi-empirical models [14]. These models were developed through curve fitting of blast parameter data obtained empirically, together with physical laws for scaling environmental conditions to enable prediction of blast wave parameters for spherical free air and hemispherical surface bursts.

Several studies have explored the use of high speed video recordings to estimate blast parameters such as explosive yield or peak overpressure [15]-[16]. In a previous work, an algorithm was developed to estimate the peak overpressure for spherical uncased charges using HSV data combined with empirical blast relationships [17]. However, while peak overpressure can be estimated through such approaches, methodologies capable of deriving additional parameters such as $t_d$ and $I_+$ directly from video data remain limited. These parameters are essential for many blast engineering applications, as impulse, in particular, is often used alongside peak overpressure as a primary indicator of structural damage potential [2].

Some previous works have explored the applicability of Kinney and Graham's hypothesis (KGH) for $t_d$ estimation in case studies [18]-[21], but without further validation. The estimation of $I_+$ for a model combining HSV and pressure gauge data using the Sadek and Gottlieb equation (SGE) [22] was also proposed in a case study, with an estimation difference below 10% at four measurement locations [23]. Moreover, an algorithm for $I_+$ estimation from HSV data only was proposed for a controlled laboratory experiment, where thorough optical and experimental calibration is required prior to testing. This method was tested against a case study experiment and resulted in an estimation difference of 15% relative to experimental measurement for a single sample [21]. However, the proposed model's application to external blast assessments is limited due to the lack of controlled optical conditions for outdoor scenarios.

Despite the availability of predictive models for blast parameters, their validation is often limited to single case studies and simple error metrics, without considering the inherent variability of experimental data conducted by different teams using different methodologies. These limitations reduce confidence in model applicability across different scenarios. The work in this paper addresses this gap by creating a robust validation framework that incorporates experimental data uncertainty across a range of explosives from a range of works to fully evaluate a hybrid optical-theoretical method for positive phase parameters estimation.

The objective of this study is to propose a bespoke methodology for estimating positive phase blast parameters for spherical uncased air bursts using the time of arrival data only. This would enable a complete blast damage assessment based solely on high speed video, without the need for pressure gauge measurements. The proposed approach combines a previously developed $t_a$-based $P_{max}$ estimation method [17], a $t_d$ estimation approach based on the KGH [1], and an $I_+$ estimation formulation derived from the SGE [22]. The performance of the methodology is evaluated using experimental datasets available in the literature involving several explosives.

## 2. Theory

This section will present the theoretical background relevant aspects for the analysis proposed in this work.

### 2.1 Scaling factors

The scaling factors used in this work account for differences in charge mass (m), ambient atmospheric pressure ($P_0$), and ambient sound speed ($a_0$) across each trial detonation [24]. The parameters are scaled according to the reference conditions: 1 kg mass, sea-level atmospheric pressure ($P_{ref}$ = 1.01325 bar), and sound speed at 15 ºC ($a_{ref}$ = 340.3 m/s). The scaling equations for pressure (P), sound speed (a), distance (R), time (t), and impulse ($I_+$) are shown in Equations (1), (2), (3), (4), and (5), respectively, where the scaled variables are denoted with an asterisk (*) after them. From this point onward, all variables are expressed in their scaled form.

$$P^* = P\left(\frac{P_{ref}}{P_0}\right) \quad (1)$$

$$a^* = a\left(\frac{a_{ref}}{a_0}\right) \quad (2)$$

$$R^* = R/\sqrt[3]{m\left(P_{ref}/P_0\right)} \quad (3)$$

$$t^* = t\,\frac{a_0}{a_{ref}}/\sqrt[3]{m\left(P_{ref}/P_0\right)} \quad (4)$$

$$I_+^* = I_+\frac{a_0}{a_{ref}}/\sqrt[3]{m/\left(P_{ref}/P_0\right)^2} \quad (5)$$

2.2 Blast wave pressure-time profile

The pressure-time profile generated by a blast wave is commonly described using the modified Friedlander equation [25] as in Equation (6). This formulation is dependent on $P_{max}$, $t_a$, $t_d$, and the wave decay constant (α). It is an empirical model which is widely accepted as an accurate representation of blast wave decay [1]. The $I_+$ can be calculated by integrating the overpressure with respect to time from $t_a$ to $t_a + t_d$, resulting in the expression given in Equation (7).

$$P^*(t^*) = P_{ref} + P^*_{max}\left(1 - \frac{t^* - t^*_a}{t^*_d}\right) e^{-\alpha\frac{t^* - t^*_a}{t^*_d}} \text{, } t_a^* \leq t \leq t_d^* \tag{6}$$

$$I^*_+ = \int_{t^*_a}^{t^*_a + t^*_d} \left[P^*(t^*) - P_{ref}\right] dt^* = P^*_{max} t^*_d \left[\frac{1}{\alpha} - \frac{1}{\alpha^2}(1 - e^{-\alpha})\right] \tag{7}$$

2.3 KB model for $P_{max}$ and the TNT equivalence

The KB models [14] are considered benchmark references due to their accuracy, particularly for $P_{max}$ predictions [26]-[28]. These semi-empirical models are based on poly-logarithmic curves fit to empirical data of each blast wave parameter. The KB models use measurements from large mass ranges scaled to 1 kg TNT through Hopinkson-Cranz scaling laws [29]-[30]. Predictions for other explosives are obtained through TNT equivalence (TNTe), typically represented by a constant that multiplies the charge mass. The scaled TNT distance (Z), from which $P_{max}$ is calculated, is related to R according to Equation (8).

$$Z = R^* / \sqrt[3]{TNTe} \tag{8}$$

Although there are some criticisms on the TNTe as a constant value for each explosive, it is not expected to change with distance after the shock wave detaches from the fireball [28].

2.4 Shock thermodynamics

The thermodynamic conditions immediately behind the shock front can be related to $P_{max}$ through the Rankine-Hugoniot relations [1]. These theoretical equations are derived from the conservation of mass, momentum, and energy across the shock front.

Initially, $P_{max}$ is predicted through the KB model according to Equation (9) where the constants $K_0$, $K_1$, $C_0$, …, $C_8$ can be found in the original report [14]. Then, the speed of sound immediately behind the shock front ($a_{max}$) can be calculated through Equation (10) [1]. Additionally, the post-shock speed of sound ($a_{ps}$) can be computed by Equation (11) [1], which represents the new local speed of sound after the air has been heated by the passage of the blast wave.

$$P_{max}^{*} = 10^{\sum_{i=0}^{8} C_i (K_0 + K_1 \log_{10} Z)^i} \quad (9)$$

$$a_{max}^{*} = a_{ref} \sqrt{\frac{P_{max}^{*}/P_{ref} + 7}{6P_{max}^{*}/P_{ref} + 7} \left(P_{max}^{*}/P_{ref} + 1\right)} \quad (10)$$

$$a_{ps}^{*} = a_{ref} \sqrt{\frac{P_{max}^{*}/P_{ref} + 7}{6P_{max}^{*}/P_{ref} + 7} \left(P_{max}^{*}/P_{ref} + 1\right)^{5/7}} \quad (11)$$

2.5 Kinney and Graham hypothesis

Kinney and Graham proposed that the end of the positive phase corresponds to the propagation of a boundary that travels approximately at the local speed of sound ($a_+$), once ambient pressure conditions are reached [1]. Based on this hypothesis, the time ($t_+$) required for the positive phase boundary to propagate from the charge radius ($R_0$) and a radial position (R) can be calculated through Equation (12), and the positive phase duration can be estimated using Equation (13).

$$t_{+}^{*}(R^{*}) = \int_{R_0^{*}}^{R^{*}} \frac{1}{a_{+}^{*}} dR^{*} \quad (12)$$

$$t_{d}^{*}(R^{*}) = t_{+}^{*}(R^{*}) - t_{a}^{*}(R^{*}) \quad (13)$$

The local speed of sound varies with the square root of the local temperature, which is assumed to reach its maximum value upon blast wave arrival and its minimum value in the post-shock region. Therefore, $a_{max}$ and $a_{ps}$ can be considered as the upper and lower bounds of $a_+$, respectively. From Equation (12), it can be observed that as $a_+$ increases, the value of $t_+$ – and consequently $t_d$ – decreases. It is therefore necessary to assess whether either yield a sufficiently accurate estimate of $t_d$, or whether a combination of both are required to achieve reliable predictions.

The local temperature rise in the near-field may be influenced not only by the blast wave itself, but also by additional detonation-related phenomena, particularly the fireball. For ideal explosives, which are the focus of this work, this additional source of thermal effects presents up to the near field boundary ($R_{nf}$), approximately 1 m/kg$^{1/3}$ [17], can be modelled as a constant $t_+(R_{nf})$ as shown in Equation (14). This result applies to any distance beyond the near field and emerges from the adopted theoretical formulation. In summary, the expected general behaviour of $t_d$ beyond the near field can be described as a combination of the KGH applied using $a_{max}$ and $a_{ps}$ as lower and upper bounds, respectively, together with a constant adjustment accounting for the additional thermal effects within the near field.

$$t_+^*(R^*) = \int_{R_0^*}^{R_{nf}^*} \frac{1}{a_+^*} dR^* + \int_{R_{nf}^*}^{R^*} \frac{1}{a_+^*} dR^* = t_+^*(R_{nf}^*) + \int_{R_{nf}^*}^{R^*} \frac{1}{a_+^*} dR^* \tag{14}$$

2.6 Sadek and Gottlieb formulation

Once the positive phase duration has been estimated, the positive phase impulse can be obtained using the theoretical formulation proposed by Sadek and Gottlieb [22]. This formulation is based on the Euler flow equations, with the flow parameters derived from the Rankine-Hugoniot relations, and the blast wave decay modelled by the modified Friedlander equation. In this work, $P_{max}$ is calculated using the KB model from Equation (9) and the pressure at the shock front arrival $P(t_a)$ is found through Equation (15). The pressure derivative with time can be calculated through Equations (16), where the blast wave speed (U) is computed from Equation (17).

$$P^*(t_a^*) = P_{max}^* + P_{ref} \tag{15}$$

$$\frac{dP^*(t_a^*)}{dt^*} = \frac{dP_{max}^*}{dt} = \left(\sum_{i=1}^{8} iC_i(K_0 + K_1 \log_{10} Z)^{i-1}\right) \frac{K_1 P_{max}^* U^*}{Z\sqrt[3]{TNTe}} \tag{16}$$

$$U^* = a_{ref}\sqrt{\frac{6}{7} P_{max}^*/P_{ref} + 1} \tag{17}$$

The SGE is given in (18), where γ, which is the ratio of specific heats, equals 1.4 beyond the near field region [1]. It relates the temporal partial derivative of the pressure behind the shock wave to the flow properties, namely the peak overpressure ($P_{max}$) and its derivative, speed of sound ($a_{max}$), particle velocity (u) and its derivative —calculated using Equation (19) and Equation (20), respectively—and the wave speed (U). The pressure partial derivative at $t_a$ can also be computed from Equation (6), with the result expressed in Equation (21). The value of α can be determined by equating the right-hand sides of Equations (18) and (21).

$$\frac{\partial P^*(t_a^*)}{\partial t^*} = \frac{[a_{max}^{*2} + u^*(U^* - u^*)]\frac{dP_{max}^*}{dt^*} + \gamma U^* P^*(t_a^*)\frac{du^*}{dt^*} + \frac{2\gamma U^* P^*(t_a^*) u^*(U^* - u^*)}{R^*}}{a_{max}^{*2} - (U^* - u^*)^2} \tag{18}$$

$$u^* = 5a_{ref}\frac{P_{max}^*/P_{ref}}{\sqrt{42P_{max}^*/P_{ref} + 49}} \tag{19}$$

$$\frac{du^*}{dt^*} = 35a_{ref}\frac{3P_{max}^*/P_{ref} + 1}{P_{ref}\left(42P_{max}^*/P_{ref} + 49\right)^{3/2}}\frac{dP_{max}^*}{dt^*} \tag{20}$$

$$\frac{\partial P^*(t_a^*)}{\partial t^*} = -\frac{P_{max}^*}{t_d^*}(1 + \alpha) \tag{21}$$

## 3. Materials and Methods

This section describes the general procedures used to generate the predictions of positive phase blast parameters. It then outlines how the validation assessment is performed using a combination of error-based and uncertainty-based criteria. This approach ensures that prediction accuracy is evaluated both in absolute terms and relative to the variability of experimental measurements.

### 3.1 Prediction algorithm description

The proposed methodology aims to estimate the positive phase duration, $t_d$, and impulse, $I_+$, which are primary descriptors of blast loading and are directly related to structural response. This is achieved using time of arrival, $t_a$, data as the primary input, which can be obtained from video recordings without the need for pressure gauge measurements. The structure of the selected models allows nonlinear relationships to be captured in a relatively simple manner.

First, the TNTe value to be used for $P_{max}$ estimation is calculated for each explosive. It is done through $t_a$ data obtained from HSV or pressure gauges, which will serve as input to a previously developed model [17]. Since the Rankine-Hugoniot relations state that the derivative of $t_a$ with respect to position is physically related to $P_{max}$ [1], this model estimates the TNTe value that best fits the experimental $t_a$ data in order to predict $P_{max}$. Then, the Z value is found through Equation (8), which is used to compute $P_{max}$ by Equation (9). Next, $P_{max}$ is used to calculate $a_{max}$ and $a_{ps}$ through Equations (10) and (11), respectively.

Subsequently, $t_+$ is estimated using the KGH for $a_{max}$ and $a_{ps}$ from the charge radius to the desired position by Equation (12), and $t_d$ is computed through Equation (13), which requires the $t_a$ data as well, and the results are analysed to evaluate how they can be combined to better fit the experimental measurements. As all points under analysis are located beyond the near field, a constant correction may be applied to account for fireball influence, as explained in Section 2. Finally, $I_+$ estimation through the SGE is performed by equating the right-hand sides of Equations (18) and (21).

To enable comparison between different explosive charges and experimental conditions, the positive phase blast parameters were scaled using the scaling relations from Equations (1) to (5) [24]. The resulting scaled parameters were then plotted as functions of scaled distance to allow comparison across different datasets.

It should be noted that the KB empirical relations [14] are used in this work solely to estimate $P_{max}$ and its derivative from the TNTe estimation. The proposed methodology does not aim to replace the KB model for peak overpressure prediction, but rather to use it to support the estimation of additional blast parameters when only video data is available.

### 3.2 Datasets for analysis

The performance assessment of the proposed methodology requires experimental datasets containing reliable measurements of blast wave parameters. The datasets used in this study were obtained from published literature on incident waves of spherical uncased charges detonated in free air which include at least measurements of $t_a$ and $t_d$ [19],[23],[31]. Only datasets containing multiple trials with multiple samples at each measurement point were considered, allowing the calculation of 95% confidence intervals (CI95). The ideal explosives included in the analysis are Bulk PE4, Cartridge PE4, PE7, Composition B, and PETN.

The $t_a$ versus R graphs for each trial of each explosive are presented in Fig. 2. The empirical mean and CI95 values of $P_{max}$, $I_+$ and $t_d$ calculated for this work from a number of samples (NS) of pressure gauge measurements for each explosive and measurement points are presented in Table 1. Bulk and cartridge PE4, and PE7 blast parameters measurements were acquired from pressure gauges and were taken according to Appendix A. It should be noted that all measurement points are located outside the near field region, $R^* > 1$ m/kg$^{1/3}$, with the exception of one PETN case situated at the near field boundary. This implies that the model can only be validated for distances beyond the near field.

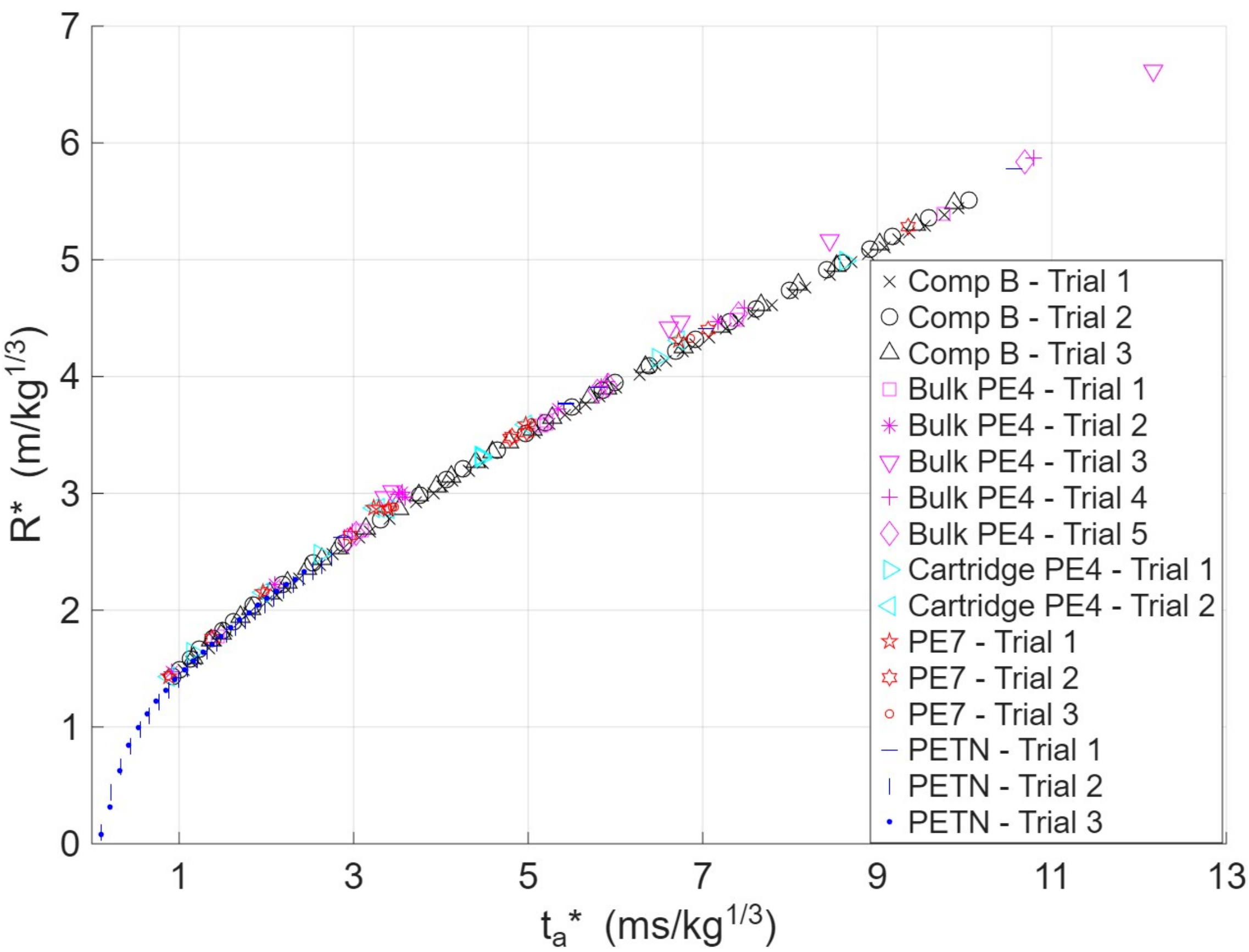

Fig. 2. $t_a$ versus R measurements. Bulk PE4, Cartridge PE4, PE7, Composition B, and PETN data. Each different marker type for the same colour represents a different trial. The datasets for the first three explosives were obtained using pressure gauges, while the remaining datasets were acquired using high speed video.

Table 1. Compilation of experimental measurements.

| Ref. | Explosive | NS | R* (m/kg$^{1/3}$) | $P_{max}$* (bar) | $I_+$* (bar.ms/kg$^{1/3}$) | $t_d$* (ms/kg$^{1/3}$) |
|---|---|---|---|---|---|---|
| [31] | Bulk PE4 | 10 | 2.63 ± 0.03 | 1.34 ± 0.05 | 0.64 ± 0.03 | 1.56 ± 0.08 |
| | | 3 | 2.99 ± 0.04 | 1.1 ± 0.1 | 0.66 ± 0.07 | 2.0 ± 0.2 |
| | | 4 | 3.6 ± 0.1 | 0.8 ± 0.1 | 0.54 ± 0.04 | 2.3 ± 0.2 |
| | | 6 | 3.92 ± 0.03 | 0.61 ± 0.02 | 0.482 ± 0.008 | 2.2 ± 0.2 |
| | | 5 | 4.53 ± 0.06 | 0.47 ± 0.03 | 0.41 ± 0.06 | 2.4 ± 0.3 |
| | | 3 | 5.86 ± 0.04 | 0.32 ± 0.02 | 0.32 ± 0.03 | 2.8 ± 0.4 |
| [31] | Cartridge | 3 | 2.87 ± 0.01 | 1.1 ± 0.1 | 0.7 ± 0.1 | 1.8 ± 0.2 |
| | PE4 | 3 | 3.31 ± 0.03 | 0.8 ± 0.1 | 0.57 ± 0.04 | 2.1 ± 0.3 |
| | | 2 | 4 ± 1 | 0.54 ± 0.04 | 0.48 ± 0.06 | 2.5 ± 0.7 |
| [31] | PE7 | 4 | 2.6 ± 0.1 | 1.4 ± 0.4 | 0.7 ± 0.1 | 1.4 ± 0.4 |
| | | 3 | 2.88 ± 0.01 | 1.1 ± 0.2 | 0.67 ± 0.09 | 1.9 ± 0.3 |
| | | 4 | 3.52 ± 0.09 | 0.77 ± 0.05 | 0.58 ± 0.04 | 2.18 ± 0.09 |
| | | 3 | 3.8 ± 0.2 | 0.65 ± 0.05 | 0.53 ± 0.04 | 2.4 ± 0.2 |
| | | 3 | 4.4 ± 0.1 | 0.51 ± 0.08 | 0.46 ± 0.07 | 2.4 ± 0.2 |
| [23] | Comp B | 2 | 2.08 ± 0.01 | 1.9 ± 0.2 | 0.82 ± 0.03 | 1.4 ± 0.1 |
| | | 2 | 2.40 ± 0.01 | 1.4 ± 0.1 | 0.74 ± 0.01 | 1.5 ± 0.2 |
| | | 2 | 2.72 ± 0.01 | 1.09 ± 0.04 | 0.69 ± 0.02 | 1.89 ± 0.07 |
| | | 2 | 3.20 ± 0.02 | 0.79 ± 0.05 | 0.59 ± 0.02 | 2.23 ± 0.08 |
| [19] | PETN | 3 | 0.99 ± 0.01 | - | - | 0.49 ± 0.09 |
| | | 2 | 1.5 ± 0.1 | - | - | 0.8 ± 0.1 |
| | | 2 | 2 ± 1 | - | - | 1.2 ± 0.4 |
| | | 3 | 2.6 ± 0.1 | - | - | 1.6 ± 0.1 |

3.3 Validation assessment criteria

The performance of the proposed models was evaluated using several complementary indicators designed to assess both the accuracy and the consistency of the estimations. For each selected metric, recent studies addressing similar scenarios assessment of blast wave parameters were consulted to define acceptance performance thresholds. The model is considered validated if it performed at least at an acceptable level across all metrics.

The relative accuracy of the model was first evaluated by comparing the mean of the estimated values with the mean of the gauges measurements. The overall predictive capability was assessed using the Mean Absolute Percentage Error (MAPE). The maximum deviation (MD) was then computed. Finally, the systematic bias was evaluated to assess the deviation of the mean prediction values relative to the mean measured values. It was calculated as the average percentage difference between the mean predictions and the corresponding mean measurements at each distance.

Regarding the first two parameters, a previous study presented a numerical model capable of predicting $t_d$ and $I_+$ for a case study with a scenario similar to that of the present work, achieving a MD of 22% and 12%, and a MAPE of 14% and 6%, respectively [32]. Another recent numerical model predicted $t_d$ and $I_+$ for a free air burst of a spherical charge, reporting a MD of approximately 40% for a case study [33]. The first one was proposed especially for the near field analysis, while the second was made for complex scenarios analysis. Moreover, an empirical model for $P_{max}$ for a similar configuration reported a MD of approximately 10% and a MAPE of approximately 5% [34]. All three models can work without any empirical input, but the first two require significant computational effort, while the last one only predicts $P_{max}$. Based on these references, MD is considered excellent up to 10% and acceptable up to 25%. MAPE is considered excellent or acceptable up to 6% and 14%, respectively. Then, the percentage bias threshold is assumed to be half of the corresponding MAPE thresholds, i.e., excellent up to 3% and acceptable up to 7%.

The agreement between the model predictions and experimental measurements was assessed by calculating the CI95 of the percentage difference between predicted and measured values. If the CI95 included 0%, the prediction was considered statistically consistent with the experimental measurement at that location. Although confidence intervals formally represent uncertainty in the mean, they are used here as an approximation of experimental variability to assess consistency between model predictions and measurements.

While an ideal 95% coverage is expected, not accounting for all sources of uncertainty can reduce the observed coverage [35]. For example, studies in automatic forecasting have shown that actual coverage can drop to around 70% [36]. Similarly, a recent application of blast prediction in complex scenarios using deep learning reported a CI95 coverage of 73% [37]. Based on the above, coverage values above 70% are considered acceptable.

### 3.4 Signal reconstruction

A significant way to qualitatively evaluate the model's performance is the positive phase reconstruction of the developed model's parameter predictions compared to experimental gauge recordings. Once the $t_a$ data is measured experimentally and the other positive phase blast wave parameters are predicted from it, the signal can be reconstructed through Equation (6). For this purpose, only the average predictions of each parameter will be used.

## 4. Results and discussion

This section presents the main results of the methodology application. A detailed example of the calculation procedure can be found in Appendix B.

### 4.1 TNTe for $t_a$ results

Initially, TNTe values derived from $t_a$ data for each explosive under analysis were computed as described in the last section. The $t_a$ versus R data for each trial was processed to determine the TNTe value that best fit each $t_a$ measurement set the best [17]. Subsequently, the resultant TNTe values from all trials for an explosive were used to calculate the mean and the CI95 values.

Table 2 presents the experimental values of the TNTe obtained from $P_{max}$ measurements and the TNTe values predicted by the methodology using $t_a$ data. These predicted values were then used to calculate $P_{max}$, which served as input for the $t_d$ and $I_+$ predictions. The predicted and experimental TNTe values show agreement within CI95 for all explosives, with differences remaining within the experimental uncertainty reported in Table 2.

Table 2. TNTe from $P_{max}$ measurements and from predictions based on $t_a$ data.

| Ref. | Explosive | Origin | Trials | TNTe exp | | TNTe pred | |
|---|---|---|---|---|---|---|---|
| | | | | Mean | CI95 | Mean | CI 95 |
| [31] | Bulk PE4 | Gauge | 5 | 1.5 | 0.2 | 1.3 | 0.1 |
| [31] | Cartridge PE4 | Gauge | 2 | 1.4 | 0.4 | 1.4 | 0.3 |
| [31] | PE7 | Gauge | 3 | 1.4 | 0.2 | 1.3 | 0.2 |
| [23] | Comp B | HSV | 3 | 1.1 | 0.1 | 1.2 | 0.1 |
| [19] | PETN | HSV | 3 | - | - | 1.4 | 0.3 |

4.2 Evaluation of models for $t_d$ estimation in free air bursts

The $t_d$ estimation obtained from the KGH using $a_{max}$ and $a_{ps}$ for free air bursts were evaluated using the results shown in Fig. 3. Following the expected pattern discussed in Section 2, the model from $a_{ps}$ overestimates $t_d$. However, the model based on $a_{max}$ showed a clear ability to predict $t_d$, without the need for blending with $a_{ps}$ or applying a constant correction. This suggests that near field non-modelled effects can be neglected and that the temperature at the end of the positive phase is approximately equal to that at shock arrival. Therefore, from this point onward, $t_d$ estimation will be performed using the $a_{max}$-based model only.

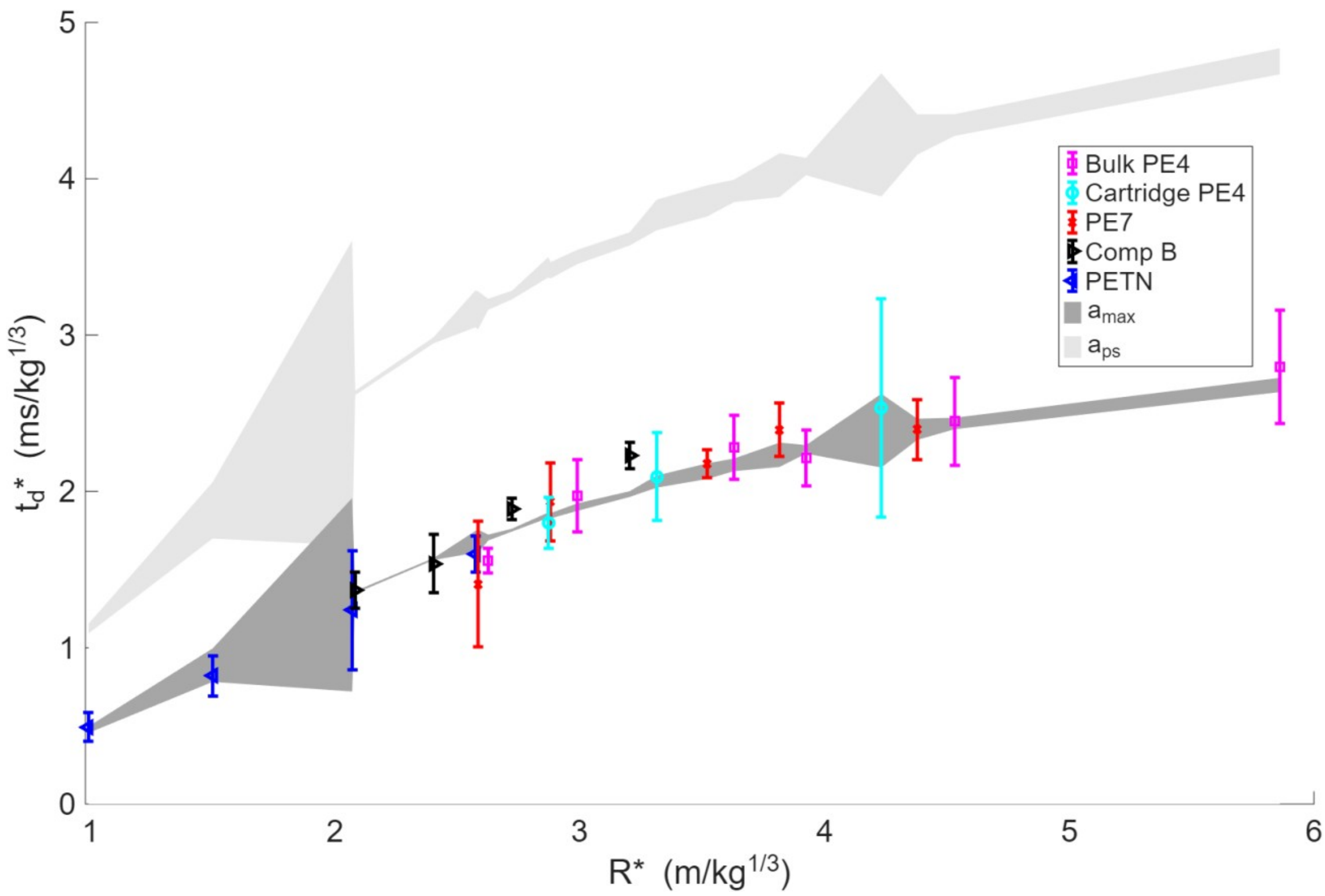

Fig. 3. Comparison of $t_d$ prediction performance for models based on $a_{max}$ and $a_{ps}$. The lighter and darker gray areas represent the CI95 predictions for $a_{ps}$ and $a_{max}$, respectively. The coloured intervals represent the mean (central value) and the CI95 boundaries (limits) of the $t_d$ measurements for each explosive.

4.3 Performance assessment: validation

Model performance was evaluated using three complementary criteria: mean and maximum prediction error, agreement with confidence intervals, and residual distribution. The mean and maximum errors are quantified using the MAPE and MD, respectively. Agreement with confidence interval is assessed through CI95 coverage. Finally, residuals distribution is evaluated using percentage bias.

An important aspect of the proposed methodology is the integration of uncertainty-based validation with traditional error metrics. By explicitly relating model errors to the variability of experimental data, the approach provides a more meaningful assessment of model performance compared to purely error-based evaluations.

Once the model for $t_d$ estimation is defined, $I_+$ can be predicted using the SGE, and the performance of both proposed approaches can be assessed. Fig. 4 and Fig.5 present the percentage differences between the estimated and the measured values of $t_d$ and $I_+$, respectively, together with the associated CI95 values for each point. Table 3 presents the performance metrics of the proposed models.

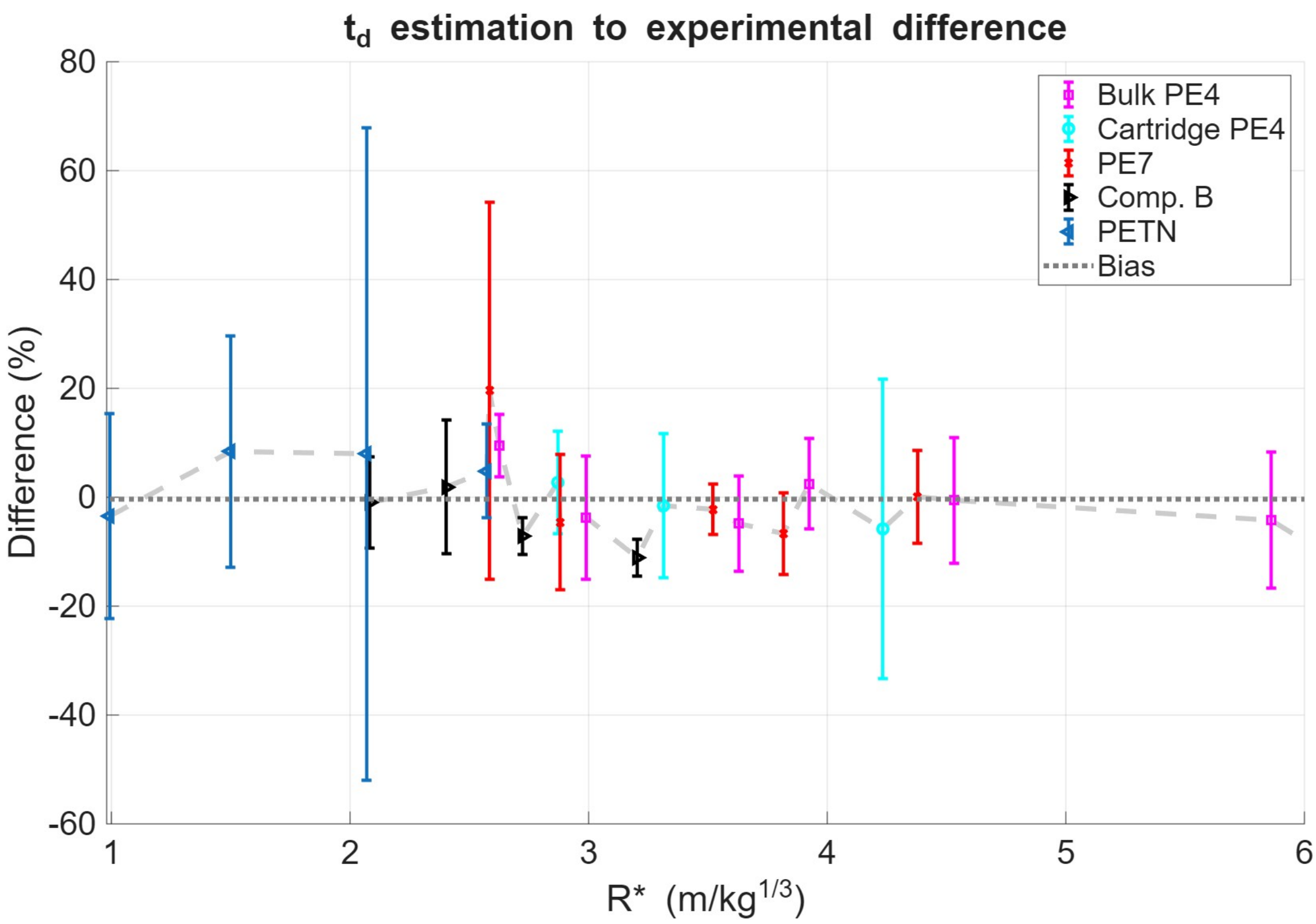

Fig. 4. Performance assessment of the $t_d$ prediction model. The close alignment between predicted and experimental values indicates good model accuracy.

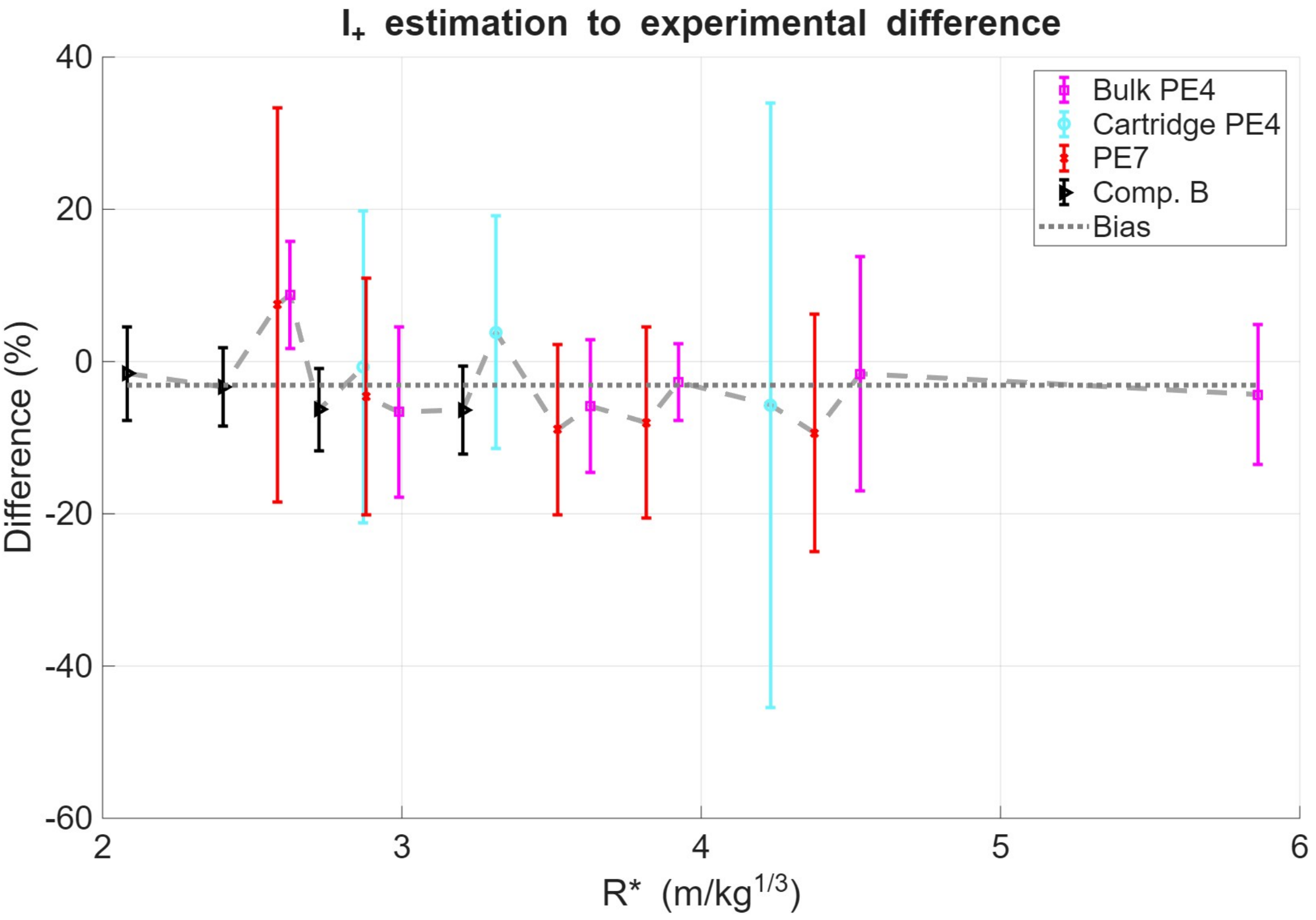

Fig.5. Performance assessment of the $I_+$ prediction model. The close alignment between predicted and experimental values indicates good model accuracy.

Table 3. Model's performance across the validation parameters metrics.

| Parameter | Metric | Criteria | Performance |
|---|---|---|---|
| $t_d$ | MAPE (%) | < 14 | 5.3 |
| | MD (%) | < 25 | 20 |
| | CI95 coverage (%) | > 70 | 86 |
| | \|Bias\| (%) | < 7 | 0 |
| $I_+$ | MAPE (%) | > 14 | 5.3 |
| | MD (%) | < 25 | 9.4 |
| | CI95 coverage (%) | > 70 | 83 |
| | \|Bias\| (%) | < 7 | 3.1 |

As shown in Table 3, the proposed models exhibit a MAPE below 6%, a MD of 20% or lower, a bias below 5%, and a CI95 coverage over 80%. The low MAPE and bias values, combined with acceptable maximum deviations and high confidence interval coverage, indicate consistent model performance across multiple validation criteria. In particular, the high CI95 coverage shows that the model prediction differences are consistent with the variability of the experimental measurements.

Deviations observed in the measurements may be due to experimental variability, which is known to be especially high for $t_d$ [27]. They may also result from the different methodologies used in the studies from which the datasets were extracted. As a semi-empirical model, the Kingery and Bulmash $P_{max}$ model [14] also contributes to the overall uncertainty.

The results in Table 3 can also be used to compare the approach proposed in this work with those of other studies. The values for the MD and MAPE for both parameters are lower than those reported for recent numerical models [32]-[33]. This indicates that, within the conditions investigated, the proposed model can predict $t_d$ and $I_+$ more accurately than the numerical approach, although its applicability is considerably narrower. Additionally, the performance of the developed model in this article is comparable to that of a semi-empirical model for $P_{max}$ [34], but offers the ability to predict all positive phase parameters with a similar level of accuracy. Therefore, the results obtained in this work indicate the competitive predictive capability of the proposed methodology compared with recently published models.

A qualitative assessment of the model performance was carried out through signal reconstitution. It was performed as explained in Subsection 3.4 and demonstrated in Fig. 6 and Fig. 7 for the positive phase pressure-time and pressure-time integral, respectively, for one sensor at each measurement point during a Cartridge PE4 charge trial [31].

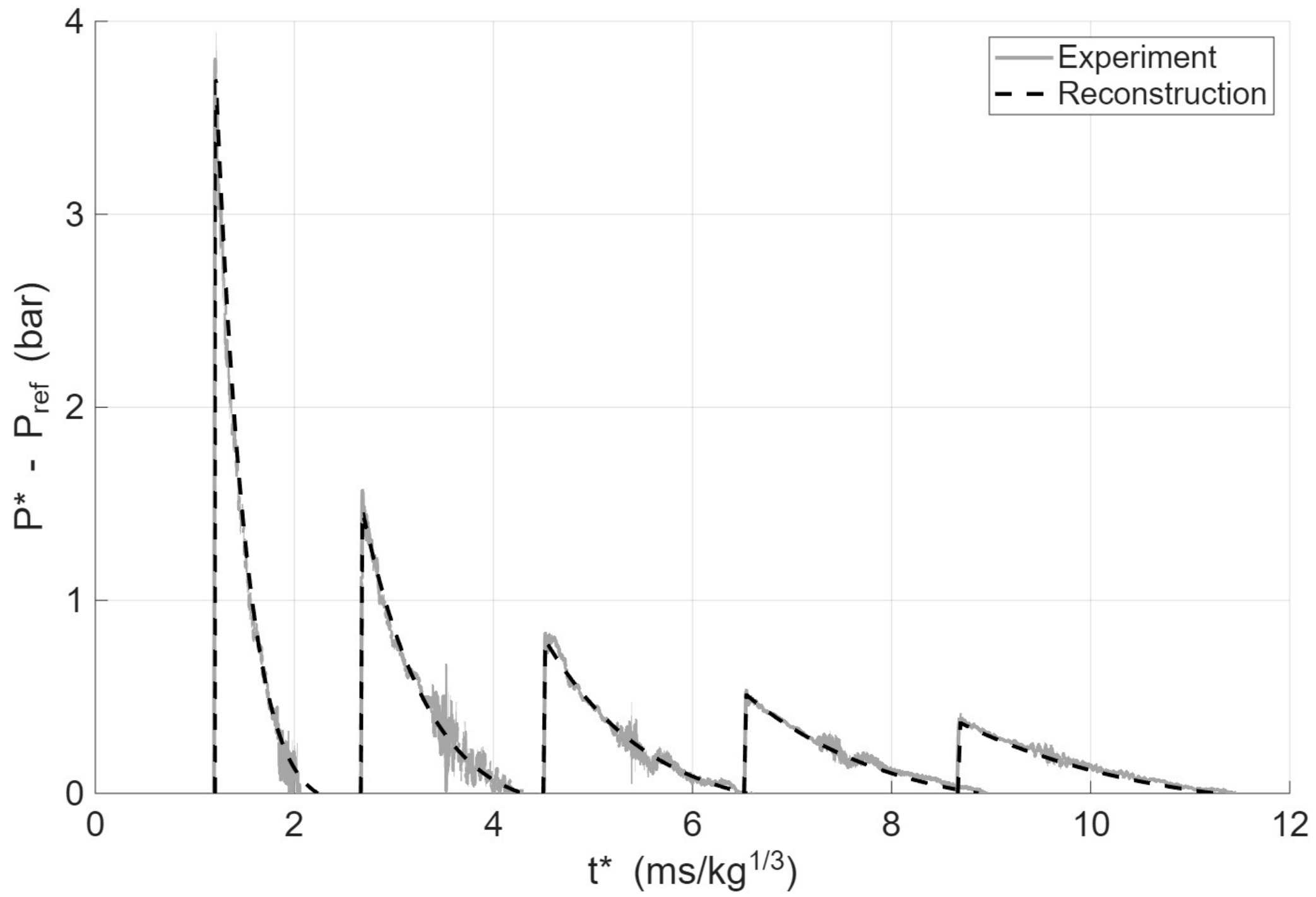

Fig. 6. Comparison between the positive phase pressure–time profiles measured by the pressure gauges and the reconstituted signals obtained from the experimental $t_a$ data and parameters predictions for one sensor at each measurement point in one Cartridge PE4 trial.

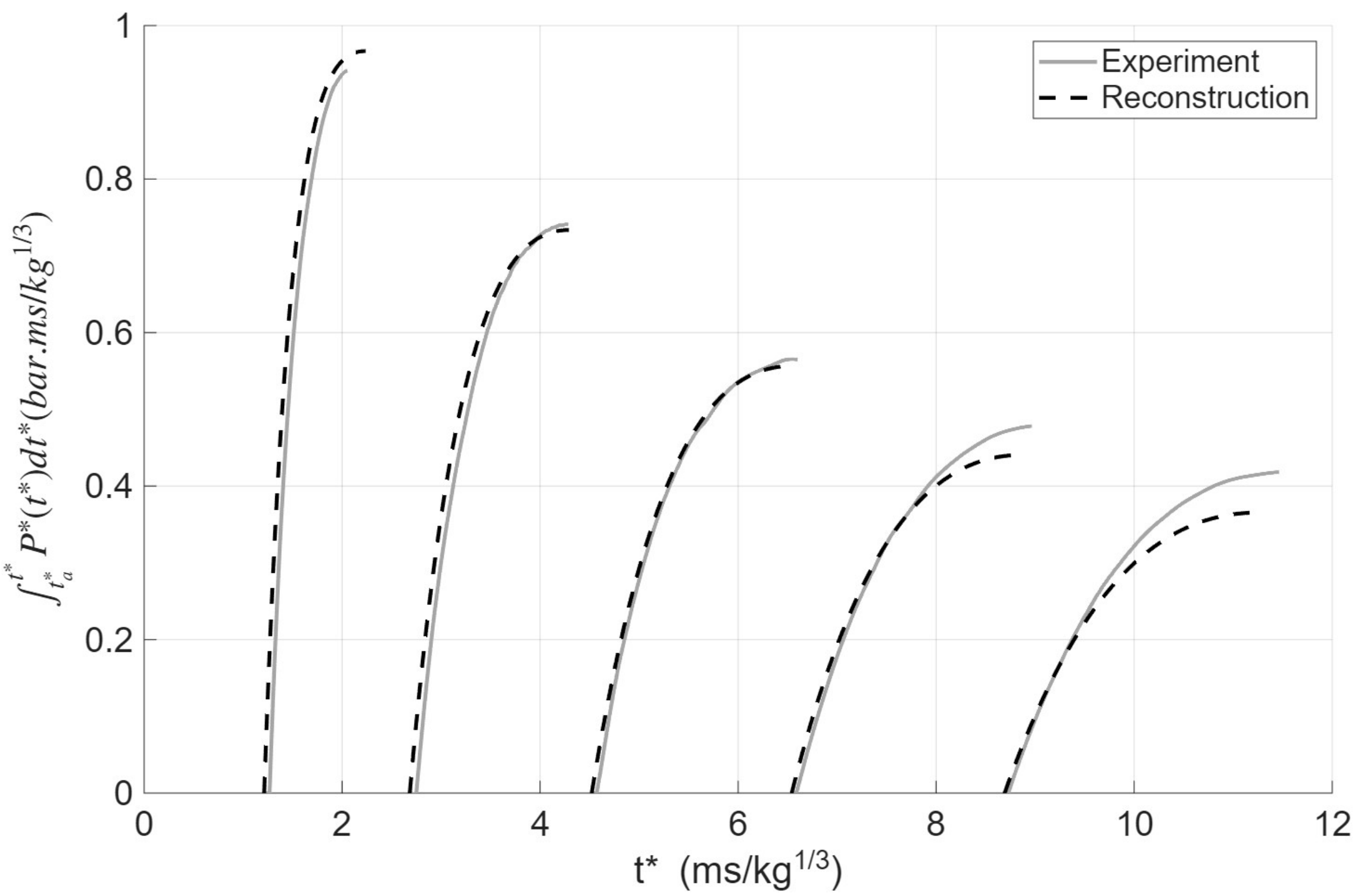

Fig. 7. Comparison of pressure-time integral profiles during the positive phase, as measured by pressure gauges, and the reconstituted signals obtained from experimental $t_a$ data and predicted parameters for one sensor at measurement point in a single Cartridge PE4 trial.

The significant agreement observed in Table 3 and Fig. 6 suggests that the developed model discussed in this article captures the dominant physical relationships governing positive phase blast parameters. The validation results obtained across multiple datasets and a wide range of parameter values indicate that the proposed model maintains consistent performance under different experimental conditions and explosive compositions.

A previous model, which required controlled optical conditions to operate, was able to predict $I_+$ in a similar scenario with a 15% deviation for a single trial in a single case study [21]. In contrast, the proposed model demonstrated comparable performance on more robust datasets, including multiple trials and multiple measurements per point for each explosive, and performed consistently across all five explosives types analysed. Importantly, it does not require controlled optical conditions.

Therefore, the predictive method for positive phase blast parameters using high speed video alone is a step changing in experimental characterisation of explosive events. The ability to gather full-field spatial loading information from a given trial reduces the need for extensive testing regimes and equipment exposure to damaging extreme loading conditions.

## 5. Conclusions

The work utilises a previously published method for extracting peak pressures from arrival time and introduces a robust and novel methodology for capturing full field spatial blast loading from high speed video using only the arrival time data. By this means, it is intended to provide an accurate way to assess the blast loading of an incident, uncased, spherical free air blast from an ideal explosive without the need for a pressure gauge. This reduces the number of trials required to perform blast loading assessments and eliminates the exposure of instrumentation to severe environmental conditions.

The methodology applies the Kinney and Graham hypothesis for positive phase duration estimation, which states that the boundary marking the end of the positive phase expands at the local speed of sound [1]. The local sound speed at this boundary is expected to lie between the shock front and post shock values for the speed of sound. It was found that the positive phase duration can be predicted on the basis of the shock front sound speed only.

The blast wave decay parameter is determined using the Sadek and Gottlieb equation [22]. This theoretical relation is based on the Rankine-Hugoniot relations, the Euler flow equations, the modified Friedlander equation [25] for the positive phase pressure-time profile, and the Kingery and Bulmash model for the peak overpressure [14].

The proposed models demonstrate strong predictive performance across all validation metrics. The positive phase duration and impulse models achieved, respectively, mean absolute percentage errors of 5.3% and 5.3%, maximum deviations of 20% and 9.4%, absolute biases of zero and 3.1%, and confidence interval coverages of 86% and 83%.

Compared with existing approaches, the proposed model shows improved or comparable accuracy. The error levels are lower than those reported for recent numerical models [32]-[33], particularly in terms of MAPE and maximum deviation, while remaining consistent with semi-empirical approaches for peak overpressure prediction [34]. In addition, compared to previous optical-based methods, which reported deviations of around 15% in controlled single-case scenarios [21], the proposed model achieves similar accuracy on more extensive multi-trial datasets without requiring controlled optical conditions.

Therefore, a novel approach to quantifying blast positive phase parameters using high speed video has been proposed with its accuracy verified against expected ranges. The validation framework provides stronger evidence that the proposed approach is suitable for engineering applications. Therefore, this work demonstrates the validity of the Kinney and Graham hypothesis for a range of explosives beyond the near field region. It also highlights the practical applicability of the Sadek and Gottlieb equation on impulse prediction.

## CRediT authorship contribution statement

**Caio B. Amorim**: Conceptualization, Methodology, Investigation, Formal analysis, Data curation, Visualization, Writing – original draft, Writing – review & editing. **Clare Knock**: Resources, Supervision, Writing – review & editing. **Dain G. Farrimond**: Writing – review & editing. **Rene F. B. Gonçalves**: Project administration, Supervision.

## Declaration of competing interest

The authors declare that they have no known competing financial interests or personal relationships that could have appeared to influence the work reported in this paper.

## Acknowledgements

The author gratefully acknowledges the support of the Brazilian Air Force and Cranfield University for enabling the research internship during which access was provided to part of the data used in this work. For the purposes of open access, the author has applied a Creative Commons Attribution (CC BY) licence to any Accepted Author Manuscript version arising from this submission.

## Acceptable Data Availability Statement (DAS)

Data supporting this study are included within the article and/or supporting materials.

Declaration of generative AI and AI-assisted technologies in the manuscript preparation process

During the preparation of this work the authors used ChatGPT (Open AI) in order to support language refinement, proofreading, and assistance in drafting and revising text based on author-provided comments and ideas. After using this tool, the authors reviewed and edited the content as needed and take full responsibility for the content of the published article.

## Appendix A: Data processing for bulk and cartridge PE4, and PE7

The measurements were extracted directly from the graphs provided in Appendix D [31], which present the pressure-time profiles in red and the time integral of pressure in blue. Fig. A.1 illustrates the procedures used to extract the measurements from the graphs.

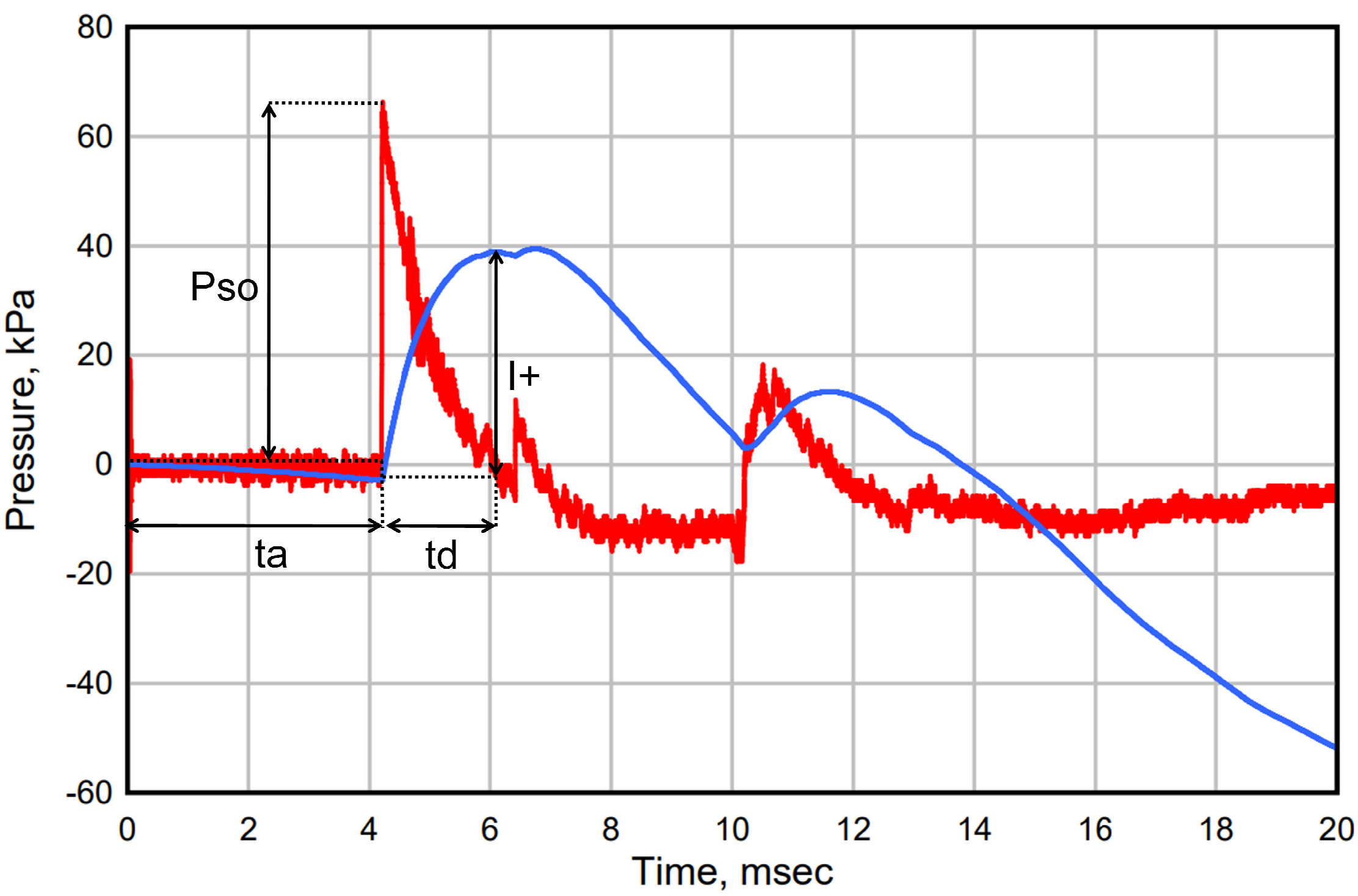

Fig. A.1. Visual depiction of how measurements were extracted from the graph generated by the pressure sensor.

## Appendix B: Methodology application example

To demonstrate the application of the proposed methodology, this section presents a step-by-step calculation for the Composition B dataset [23]. The dataset was chosen especially because of the availability of high speed video data.

The first step is to assess the best TNTe value to fit the $t_a$ versus R data provided by the high speed video. This is assessed using the KB model for $P_{max}$ for spherical free air bursts [14], adapted to predict $t_a$ using the Rankine-Hugoniot relations [1] through a previously published model [17]. The TNTe values for the three trials presented in the study are 1.12 ± 0.01, 1.16 ± 0.01, and 1.20 ± 0.02, resulting in a final TNTe value of 1.2 ± 0.1. The final uncertainty reflects both the variability between the three trials and the individual measurement uncertainties, combined using standard uncertainty propagation.

The first pressure gauge data acquisition point for Composition B shown in Table 1 at R* = 2.08 ± 0.01 $m/kg^{1/3}$, was chosen to demonstrate the next steps. For simplicity, the $t_a$ value used in the methodology was predicted using the calculated TNTe value [17] instead of being taken directly from the measurements, especially because the high speed video data may not provide a $t_a$ measurement for all trials for a specific point. Table B.1 presents the $t_d$ and $I_+$ calculation methodology by showing the intermediate values computed and the equations used for this. For simplicity only the mean values of the variables are shown.

Table B.1. Methodology calculation steps.

| Parameter | Value | Calculation | Function of |
|---|---|---|---|
| $a_{ref}$ (m/s) | 340.3 | - | - |
| $P_{ref}$ (bar) | 1.01325 | - | - |
| R* (m/kg$^{1/3}$) | 2.04 | - | - |
| Z (m/ kg$^{1/3}$) | 1.96 | Equation (8) | R*, TNTe |
| $P_{max}$* (bar) | 2.04 | Equation (9) | Z |
| P*($t_a$*) (bar) | 3.05 | Equation (15) | $P_{max}$* |
| $a_{max}$* (m/s) | 406 | Equation (10) | $P_{max}$* |
| $t_+$* (s/kg$^{1/3}$) | $3.33.10^{-3}$ | Equation (12) | $a_{max}$* from $R_0$* to R* |
| $t_a$* (s/kg$^{1/3}$) | $1.98.10^{-3}$ | [17] | TNTe [(i)] |
| $t_d$*(s/kg$^{1/3}$) | $1.35.10^{-3}$ | Equation (13) | $t_+$*, $t_a$* |
| U* (m/s) | 561 | Equation (17) | $P_{max}$* |
| d$P_{max}$*/dt* (bar. kg$^{1/3}$/s) | $-1.23.10^{-3}$ | Equation (16) | $P_{max}$*, Z, U* |
| u* (m/s) | 297 | Equation (19) | $P_{max}$* |
| du*/dt* (m. kg$^{1/3}$/s$^2$) | $-1.22.10^{5}$ | Equation (20) | $P_{max}$*, d$P_{max}$*/dt* |
| ∂P*/∂t* (bar.kg$^{1/3}$/s) | $-4.36.10^{3}$ | Equation (18) | $P_{max}$*, U*, d$P_{max}$*/dt*, u*, du*/dt*, R* |
| α | 1.89 | Equation (21) | $P_{max}$*, $t_d$*, ∂P*/∂t* |
| $I_+$* (bar.s/ kg$^{1/3}$) | $8.04.10^{-4}$ | Equation (7) | $P_{max}$*, $t_d$*, α |

[(i)] $t_a$ can also be obtained through interpolation of the measurements.